# Efficient labeling of solar flux evolution videos by a deep learning model

Subhamoy Chatterjee [1,*,+], Andrés Muñoz-Jaramillo [1,+], and Derek A. Lamb [1]
[1] Southwest Research Institute, Boulder, CO 80302, USA
[+] these authors contributed equally to this work
* subhamoy@boulder.swri.edu

**ABSTRACT**
Machine learning (ML) is becoming a critical tool for interrogation of large complex data. Labeling, defined as the process of adding meaningful annotations, is a crucial step of supervised ML. However, labeling datasets is time consuming. Here we show that convolutional neural networks (CNNs), trained on crudely labeled astronomical videos, can be leveraged to improve the quality of data labeling and reduce the need for human intervention. We use videos of the solar magnetic field, crudely labeled into two classes: emergence or non-emergence of bipolar magnetic regions (BMRs), based on their first detection on the solar disk. We train CNNs using crude labels, manually verify, correct labeling vs. CNN disagreements, and repeat this process until convergence. Traditionally, flux emergence labelling is done manually. We find that a high-quality labeled dataset, derived through this iterative process, reduces the necessary manual verification by 50%. Furthermore, by gradually masking the videos and looking for maximum change in CNN inference, we locate BMR emergence time without retraining the CNN. This demonstrates the versatility of CNNs for simplifying the challenging task of labeling complex dynamic events.

## Introduction

Big-data problems have become increasingly common in astronomy [1]. Large datasets present complex challenges that cannot be tackled with traditional computational techniques. Supervised machine learning (ML), e.g. deep learning, is a promising and effective technique for the classification of complex data such as images and videos [2]. However, manually labelling large databases is a laborious process that requires time and consistency. Iterative labelling approaches, such as 'active learning' [3] can significantly save time, reducing the cost of making big-data ML ready.

'Bipolar flux emergence', which involves the appearance of bipolar/complex magnetic regions (BMRs) on the solar surface, is an example of complex dynamical events that are difficult to label. These regions have the potential to drive space weather events such as flares and coronal mass ejections that can negatively affect satellite networks and long-distance communication [4–6]. Techniques for detecting flux emergence and interactions typically rely on image-by-image processing to perform segmentation and tracking of magnetic elements for detecting dynamic phenomena such as appearance, disappearance, splitting and merging [7–11]. Recently, the single-image segmentation component of this process has been performed using CNNs [12]. However, the complex interaction of magnetic elements demands increasingly complex tracking codes and limits the possibility of real-time detection. For this reason, we have created an end-to-end deep-learning approach for classifying videos of magnetic patch evolution without explicitly supplying segmented images, tracking algorithms or other hand-crafted features. The idea is to allow relevant information abstraction to autonomously take place deeper in the CNN layers to detect flux elements and describe their interaction towards appropriate classification. We start with a crudely labelled solar flux emergence dataset, train a deep learning model, refine the labelling using the trained model and, finally, show how a simple change in data input allows us to detect the time of emergence.

## Results

We use 96-minute cadence videos of SoHO/MDI [13] Line-of-Sight (LoS) magnetic patches (15° × 15° in Carrington grid) (Figure 1) labelled for BMR 'emergence' (i.e. BMRs that clearly emerge within the visible solar disk) or BMR 'non-emergence' (i.e. BMRs that rotate into view). We selected this patch extent as we wanted an integer number of non-overlapping boxes per 360° longitude and 180° latitude. Also, the extent of 15° was large enough to encompass all but the largest few BMRs from the BARD catalog. Again, since we are interested in the emergence process, the final size of the BMR need not set an upper limit on the size of our box, as long as the box size is sufficient to capture the emergence. We use BMRs from the Bi-polar Active Region Detection (BARD) code [14] to produce the initial labels. Our initial emergence vs. non-emergence labels are crude in nature, based simply on the first observation of each BMR in the catalog. Any BMR whose first observation occurs to the west of -60° heliographic longitude, with respect to the solar central meridian, is labeled as an emergence and the rest of as non-emergence. We use the following steps to bring uniformity in the videos as input to a classification model:
1. Consecutive frames in the videos are separated in time by 96 minutes, i.e. the cadence of MDI full-disc magnetograms. However, data-gaps of MDI cause frame gaps in our videos. Based on the difference in time-stamps of the consecutive video frames we identify missing frames. As CNN is blind to the timestamps of individual frames, we need to make sure that every two consecutive frames have a constant time difference i.e. 96 minutes. For this purpose we perform interpolation. As the frames are projected to the Carrington grid, we linearly interpolate missing frame pixels using available frame pixels over the time axis for all the locations across the frames.
2. Regions outside the solar limb are replaced with a background field of 0 Gauss
3. Videos are padded with 0 Gauss frames towards the end to make all videos have a standard duration of 225 frames.

This translates to a video classification problem with inputs of size 90×90 pixels (in a Carrington longitude-latitude grid), 225 time frames ($\gtrsim$ 1/2 solar rotation), and output either 1 (emergence) or 0 (non-emergence). This is to be noted, that MDI pixels subtend 2", and this corresponds to approximately 1450 km on the solar surface, at the disk center. From the opposite perspective, 1 heliocentric degree subtends approximately 12130 km on the solar surface. The ratio of these i.e. (12130 km/heliocentric degree) / (1450 km at disc center / MDI pixel) = 8.36 MDI pixels / heliocentric degree at disk center, gives the Carrington plate scale that would preserve the MDI resolution at the disk center. Since 1) most BMRs will not appear at disk center and 2) we are not interested in the smallest resolvable regions anyway, we chose a slightly courser sampling of 6 pixels per heliocentric degree (i.e. 90 pixels/15°) for the Carrington maps.

To help the CNN with the training process we normalize the input video frames (*im*) first by clipping the fields within [-1000 Gauss, 1000 Gauss] and then applying the transformation $\frac{1}{2}(1 + \frac{im}{1000})$. Thus 0, 0.5 and 1 in the normalised frames represent -1000 Gauss, 0 Gauss and 1000 Gauss fields respectively.

We divide our data into a training+validation set (BMR observations within the first 10 months of every year between 1996 and 2011; Figure 2) and a test set (BMR observations on the last two months of every year between 1996 and 2011; Figure 2). We exclusively use the training+validation set to train the CNN weights and optimize CNN hyper-parameters, and set aside the test set to evaluate the performance of our algorithm under pseudo-operational conditions (i.e. with data that the CNN has never seen). We make all our design decisions based on the training+validation set. The test set is only used to assess the final performance shown this article.

For every model we augment each video with vertical and polarity flips resulting in an increase of the number of our training and validation samples by a factor of 4. This is justified because the characteristics of magnetic flux emergence are not expected to be different based on whether they occur in the northern or southern hemisphere of the Sun. Similarly, hemispheric polarity orientation changes sign with every new solar cycle without this having any measurable impact on the properties of BMRs.

We use a CNN [15] based on the Visual Geometry Group (VGG) architecture [16]. This architecture performs information abstraction with repeated convolution layers followed by a non-linear activation called 'swish' activation [swish(x) $= \frac{x}{1+e^{-x}}$][17]. After each convolutional layer we perform max-pooling, gradually reducing the patch size and increasing the number of channels ending up in 1D vectors (Figure 1). Subsequently, the fully-connected layers at the end (Figure 1) generate the classification outcome with a sigmoid activation [18]. To regularize the model we add drop-outs [19] with a rate of 0.5 in the last two fully connected layers before the output layer. To optimize the network parameters during training we use a 'stochastic gradient descent' [20] optimizer with a learning rate of $10^{-4}$, momentum of 0.9 and a batch size of 10.

## *Iterative Relabelling: Reducing Manual Intervention*

For efficiently relabeling the data, we split the training+validation set (a total of 2032 videos) into 5 blocks of two months each and use the blocks to assemble five training+validation permutations to train 5 different models (models 1-5; Figure 2). This way, each one of these two-month blocks has an associated model where this block is not used to train CNN weights. This enables the unbiased evaluation of the classification of all videos and an assessment of the quality of the crude initial labeling.

We train 5 models and evaluate them on their respective 5 non-overlapping validation blocks (Figure 2). Subsequently, we manually check and relabel (if necessary) the videos where CNN disagrees with the crude labeling (both false positives and false negatives). During each pass, we re-train all models with the verified data and repeat this verification and relabeling process until convergence (see detailed algorithm in 'Methods' section).

Figure 3 shows the relabelling process through a bar chart. We find that for higher passes an increasing fraction of the CNN labeling "mistakes" (false positives and negatives) are actually mistakes in the initial crude BARD labeling. After convergence, we manually check all the videos that the models classified successfully and find that ∼90% of them had been correctly labelled through this process, even though we had only manually verified ∼50% of them. This translates into a significant reduction of human effort when working with larger datasets.

## *Model-ensemble Training and Performance Evaluation*

We quantify the uncertainty in emergence/non-emergence classification using a different ensemble of the trained models (models 6-9; Figure 2). In this case we set aside months 9 and 10 as a common benchmark ('benchmarking set'; Figure 2) and use the remaining four blocks to train and validate four permutation models like those described in the previous section. Additionally, we create two random realization per model by changing the random seed that determines the order in which the neural network sees the videos and the drop-out (random removal) of neuron connections.

We use the median of the final sigmoid activation output of these 8 models (model 6-9 × 2 random realizations) as the metric to classify each video. We find that an optimal threshold of 0.6 applied to the median of sigmoid model output yields the True Positive (TP) maximum accuracy. We derive this threshold from the intersection point of precision ($\frac{True\,Positive\,(TP)}{True\,Positive\,(TP) + False\,Positive\,(FP)}$) and recall ($\frac{True\,Positive\,(TP)}{True\,Positive\,(TP) + False\,Negative\,(FN)}$) curves (Figure 4). Finally, we add one more random realization for each of models 1-5 and use the tuned 0.6 probability threshold (derived from models 6-9) to classify the 'test set' and assess performance under pseudo-operational conditions. Using model ensemble median as the classification metric, we achieve 83% classification accuracy. We also evaluate other binary classification performance metrics such as True Skill Statistic (TSS) and Matthews Correlation Coefficient (MCC) defined by: $TSS = \frac{TP}{TP+FN} + \frac{TN}{TN+FP} - 1$ and $MCC = \frac{TP \times TN - FP \times FN}{\sqrt{(TP+FP)(TP+FN)(TN+FP)(TN+FN)}}$. We find both TSS and MCC to be 0.65. Figure 4 shows model inferences on the 'test set' as 2D histograms over videos and emergence probability, with the histogram frequency being the number of models that output a certain probability value from sigmoid activation. We divide the 2D histograms into two regions, which we name unstable and stable inferences for the purpose of uncertainty estimation in classification accuracy. Stable (unstable) inferences encompass the videos for which the probability threshold of 0.6 lies outside (inside) the model-ensemble's $25^{th}$-$75^{th}$ percentile range. We define unstable classifications as those that will change if another percentile is used as reference. As observed in Figure 4, the ensemble of CNNs is more confident in the classification of emergence than of non-emergence, i.e. there is a larger (smaller) proportion of unstable true negatives (true positives). Because
of this, performance is more sensitive to changes in the classification of non-emergences than of emergences. We find that accuracy falls to 79% (71%) when the ensemble's $25^{th}$ ($75^{th}$) percentile is used as the classification metric instead of the ensemble median. False positives and negatives have similar ratios of stable vs. unstable videos. Thus, they do not seem to drive performance changes as much. To examine the model performance at different phases of solar cycle, we break the test into blocks of 4 months (every 2 years) and evaluate the classification accuracy on each of those blocks. As shown in bottom-panel of Figure 4, we don't find any systematic effect introduced by the solar cycle phase on accuracy.

## *Model Repurposing: Automatic Detection of Emergence Time*

As a sanity check, we test the ability of the network to autonomously determine the time of each BMR emergence. For this, we use all videos that the model ensemble classifies correctly. The idea is to identify what are the relevant frames that lead the ensemble of models to classify a video as a BMR emergence. We mask-out portions of each video, after an arbitrary number of frames, by repeating said frame until the end of the video. This is done iteratively for each video starting with the last frame, then with the last two frames, then with the last three frames, then with the last n frames, and so on, until the entire video has been masked out. We find there is a period in time where the majority of the models transition from classifying the video as an emergence into a non-emergence. The frame when that happens coincides well with the moment where the BMR emerges on the solar disk.

We quantify this by taking the time derivative of the sigmoid probability for each model in the ensemble and finding the frame with maximum gradient. The median of the frame number for all models in the ensemble is what we determine to be the frame of emergence. This process is shown in the upper (a-d) and lower panels (e-h) of Figure 5. Figure 5a and 5e depict the model-ensemble curves for one of the videos (frames shown in Figure 1). The ground-truth frame and predicted frame locations are shown with vertical dashed line and red star respectively. We also show other sample curves for false positives and true and false negatives. We note that the ensemble median at the end of the probability curve (i.e. no masked frames) is what actually determines the classification (emergence/non-emergence), depending on whether the ensemble median is greater or less than 0.6 (red dashed line in Figure 5). Figure 6 shows the time labeling of the BMR emergence, where frame color indicates the number of models classifying the video as an emergence given that all subsequent frames have been masked-out.

Figure 7 shows the accuracy of our emergence frame detection. Upper part of Figure 7 shows a scatterplot between the real emergence time and identified time of emrgence through model-ensemble. Lower part of Figure 7 shows a 2D histogram with the difference between the estimated and real time of emergence for all test set videos that the model-ensemble correctly classifies as BMR emergence. The median of all model ensembles is +0.4 days. This means that using the median of the location of maximum probability gradient tends to result in an estimation that is slightly behind the observed emergence frame. The $5^{th}$ percentile is -2.1 days, the $25^{th}$ percentile +0.1 days, the $75^{th}$ percentile +1.1 days, and the $95^{th}$ percentile +2.7 days. This means that 50% (90%) of emergence detections are within [0.1,1.1] ([-2.1, 2.7]) days of the observed emergence. For reference, a $15° \times 15°$ patch is visible on disk for $\sim 14$ days.

We note that we have not fine-tuned or optimized in any way what are the optimal detection thresholds that would to maximize accuracy of frame detection. This is beyond the scope of this paper and needs to be done carefully on the training+validation set before in can be tested on a pseudo-operational setting with the test set. Instead, our goal is to showcase the versatility and potential of CNNs. We want to highlight how training them for something as simple as a 'yes/no' question can be easily re-purposed for a much more sophisticated question like a 'yes/no and, if yes, when?'.

## Discussion

We show that a deep learning model can be harnessed to refine the initially crude labelling of the dataset used for training the model. We achieve this by partitioning the training dataset into blocks and training several models with complementary validation sets. We find that, as we progress through this iterative process, any false positive and false negatives reported by the CNN are increasingly likely to be mislabels. This iterative process lessens the manual effort needed for painstakingly labeling data by 50%, which is one of the main obstacles of deep learning applications to classification on large astronomical databases.We factored in the associate uncertainty of using a relatively small dataset by training an ensemble of models and calculating the median and range of possible classification outcomes. Using the ensemble median yields a classification accuracy of 83% of BMR flux emergences vs. non-emergences. Choosing the $25^{th}$ ($75^{th}$) percentile, instead of the median, translates into a performance reduction to 79% (71%). This asymmetry is a consequence of the CNN ensemble being more confident about classifying emergences vs. non-emergences.

Even though our model was solely trained as an emergence classifier, we show that it can be re-purposed to also detect the time of emergence by progressively freezing video frames until the emergence event is masked. We see this as evidence that our model is being able to learn, abstract, and generalize the characteristics that make a BMR flux emergence. This has interesting implications for the labeling, and classification of dynamical astrophysical events in which the exact time of the event is unknown. This implies that it may be possible to use the deep learning model for the prediction of emergence by only looking at the early evolution prior to flux emergence. Early detection of magnetic flux emergence, if possible before there are signatures visible to a human observer, is a holy grail in space weather forecasting. Coupled with a early-warning observatory placed at the L1 Lagrangian point, it would significantly increase our readiness and ability to mitigate its impact.

## Methods

**BARD.** The Bipolar Active Region Detection (BARD) catalog[14] uses a semi-automatic segmentation algorithm, coupled with human supervision to detect and track BMRs as they emerge and/or rotate in and out of view at a cadence of one observation per day.

**VGG.** This is a CNN architecture available in two varieties- VGG16 and VGG19 named after the depth of layers 16 . The convolutional layers uses 3×3 kernels. The volume is controlled by maxpooling as the depth increases. The network ends with two fully-connected layers. It is widely used for natural image classification. Our CNN architecture is based on VGG with input being videos instead of images.

**Sigmoid Activation.** The sigmoid function[18] ($sigmoid(x) = \frac{1}{1+e^{-x}}$) outputs a number within 0-1 and is typically used to represent classification probability. We use sigmoid activation at the end to evaluate the probability of a video to be an 'emergence'.

**Swish Activation.** Swish[17] a smooth activation function developed by Google to replace the widely used activation function ReLU (max(0, x)). Swish activation is represented as *swish(x) = x.sigmoid(x)* We use 'swish' activation as the non-linearity
after every convolution layer.

**Relabelling Algorithm.** The detailed steps of our iterative relabelling algorithm are shown below-
1. Divide crudely labelled data into 5 train-validation combination where $Months_{val}$ = *{2m − 1, 2m}* and $Months_{train}$ = *{1, 2, .., 10}\\Months_{val}* ∀ m ∈ *{1, 2, 3, 4, 5}*
2. Train the model (CNN) for each combination until validation accuracy reaches maximum.
3. Using the trained model inference identify false-postives (*fp*) and false-negatives (*fn*).
4. Look at each *fp, fn* and identify model mistakes (*E*) where crude labels are found to be correct
5. Calculate the mistakes of crude label |*fp*| + |*fn*| − |*E*| and change those labels
6. Calculate the ratio $R = \frac{|E|}{|fp|+|fn|-|E|}$
7. n = 1
8. $Data_{Pass[n]}$ = relabelled data
9. While R>0.5 do
    a) Divide $Data_{Pass[n]}$ into 5 train-validation combination where $Months_{val}$ = *{2m − 1, 2m}* and $Months_{train}$ = *{1, 2, .., 10}\\Months_{val}* ∀ m ∈ *{1, 2, 3, 4, 5}*
    b) Train the model for each combination until validation accuracy reaches maximum.
    c) Using the trained model inference identify new (not seen in prior passes) false-postives (*fp*) and false-negatives (*fn*).
    d) Look at each *fp, fn* and identify model mistakes (*E*) where crude labels are found to be correct
    e) Calculate the mistakes of crude label |*fp*| + |*fn*| − |*E*| and change those labels
    f) Calculate the ratio $R = \frac{|E|}{|fp|+|fn|-|E|}$
    g) n = n + 1

       h) $Data_{Pass[n]}$ = relabelled data
10. return $Data_{Pass[n]}$ , model

# Data availability
The SoHO/MDI magnetograms, used to create the flux emergence videos for this study, are available from the Joint Science Operations Center (http://jsoc.stanford.edu). All the flux evolution videos with their 'emergence', 'non-emergence' labels can be accessed through Harvard Dataverse (https://doi.org/10.7910/DVN/6F25MG)..

# Code availability
The iterative relabelling algorithm has been explicitly depicted in the 'Methods' section. The code for data preparation and training the CNN can be accessed in the form of a python notebook through https://github.com/subhamoysgit/flux_emergence/.

# Acknowledgements
This research was funded by NASA grants 80NSSC19M0165 and 80NSSC18K0671.

# Author contributions statement
S. Chatterjee and A. Muñoz-Jaramillo planned the experiments and wrote the paper.
S. Chatterjee set up and ran the experiments.
A. Muñoz-Jaramillo provided the list of events that was analyzed.
D. A. Lamb assembled the video sequences used in this work and helped edit the manuscript.

# Competing interests
The authors declare no competing interests.

# Figures

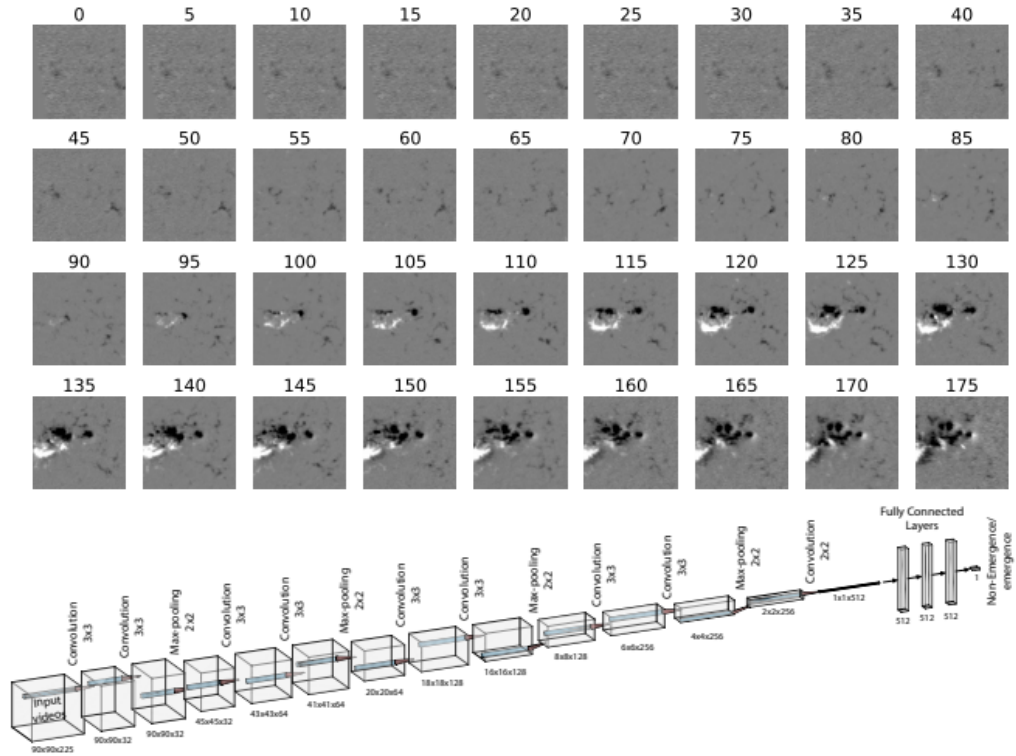

**Figure 1.** Dataset and model architecture. Upper part of the figure shows every 5th frame of an emergence video starting on November 29, 1998 17:36 hours (UT). Lower part of the figure shows adapted deep learning model for classifying videos into 'emergence' or 'non-emergence' classes. We use repeated convolution, max-pooling and fully connected layers. The operations (convolution and max-pooling with n × n kernels) are placed between the boxes (width×height×n_channels), signifying boxes as the outcomes of those operations.

| Months | Iterative Data Relabelling ||||| Uncertainty Estimation ||||
|---|---|---|---|---|---|---|---|---|---|
| | Performance Evaluation ||||| Detection Threshold Definition ||||
| | model1 | model2 | model3 | model4 | model5 | model6 | model7 | model8 | model9 |
| Jan | Training Set |||| Validation | Training Set ||| Validation |
| Feb | | | | | | | | | |
| Mar | | | | Validation | Training Set | | | Validation | Training Set |
| Apr | | | | | | | | | |
| May | | | Validation | | | | Validation | | |
| Jun | | | | | | | | | |
| Jul | | Validation | | | | Validation | | | |
| Aug | | | | | | | | | |
| Sep | Validation | | | | | Benchmarking Set |||| 
| Oct | | | | | | | | | |
| Nov | Test Set ||||| | | | |
| Dec | | | | | | | | | |

**Figure 2.** Data management used to build, train, validate and test different models for iterative relabelling, detection threshold estimation and performance evaluation. We split the data according to the months of the year to minimize the impact of the large temporal coherence present in solar data. The randomly-sampled splits typically used in other ML applications tend to overestimate performance in the solar case due to the fact that data close to each other in time can be generated by virtually the same structures.

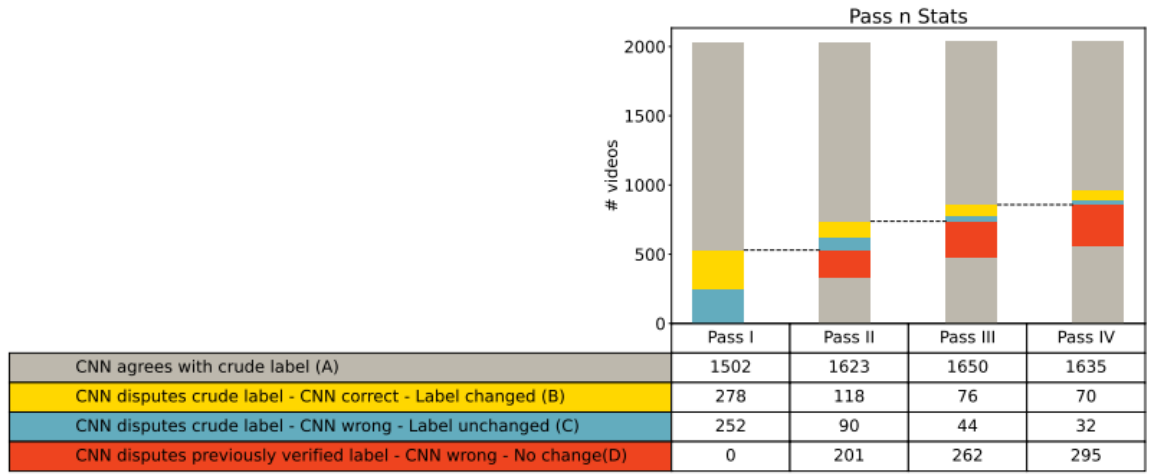

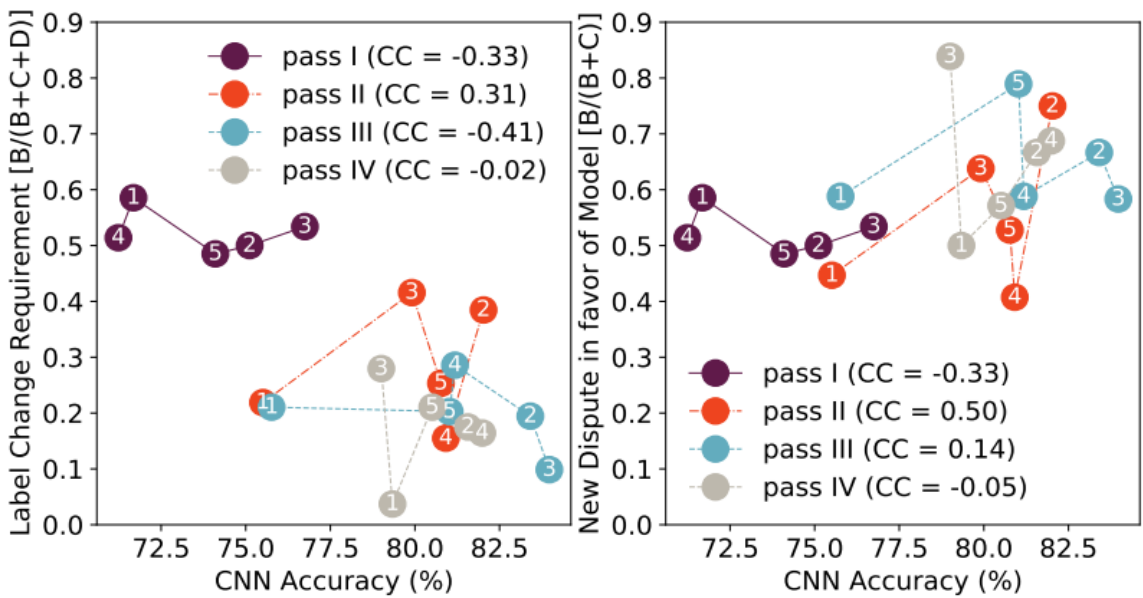

**Figure 3.** Sequence of iterative relabelling and convergence of performance. Top panel shows how the relabelling progresses through a total of 2032 videos (height of the bars) from Pass I to Pass IV. The videos are divided to 5 non-overlapping validation blocks (numbered circles in bottom panels) and are manually checked in a particular pass if the model (CNN) inference differs from the initial crude labelling and was not checked in prior passes. The red bars show the proportion of disagreements that have been manually verified in previous passes. The dashed black line shows the total proportion of videos that have been manually checked. The yellow (real mislabels) and blue (CNN mistakes) bars show newly discovered disagreements. The number of total disagreements reduces with each pass. The proportion of newly discovered disagreements that lead to a relabel increases with each pass. These trends result from labelled data makes more sense, allowing the CNN to better generalize. We quantify these trends as fractions $\frac{B}{B+C+D}, \frac{B}{B+C}$, and plot them against CNN accuracy for different validation blocks in the bottom panels. Bottom left panel captures the trend in the bar chart and shows that the need of relabeling goes down from Pass I to Pass IV, because each pass is progressively downward in the plot. Bottom right panel shows that the models tend to capture the problems in the crude labeling better as we go higher in the passes because each pass is progressively upward in the plot. Both label change requirement and correctness of the CNN in disagreement become linearly independent of CNN accuracy for higher passes, as shown by the correlation coefficients.

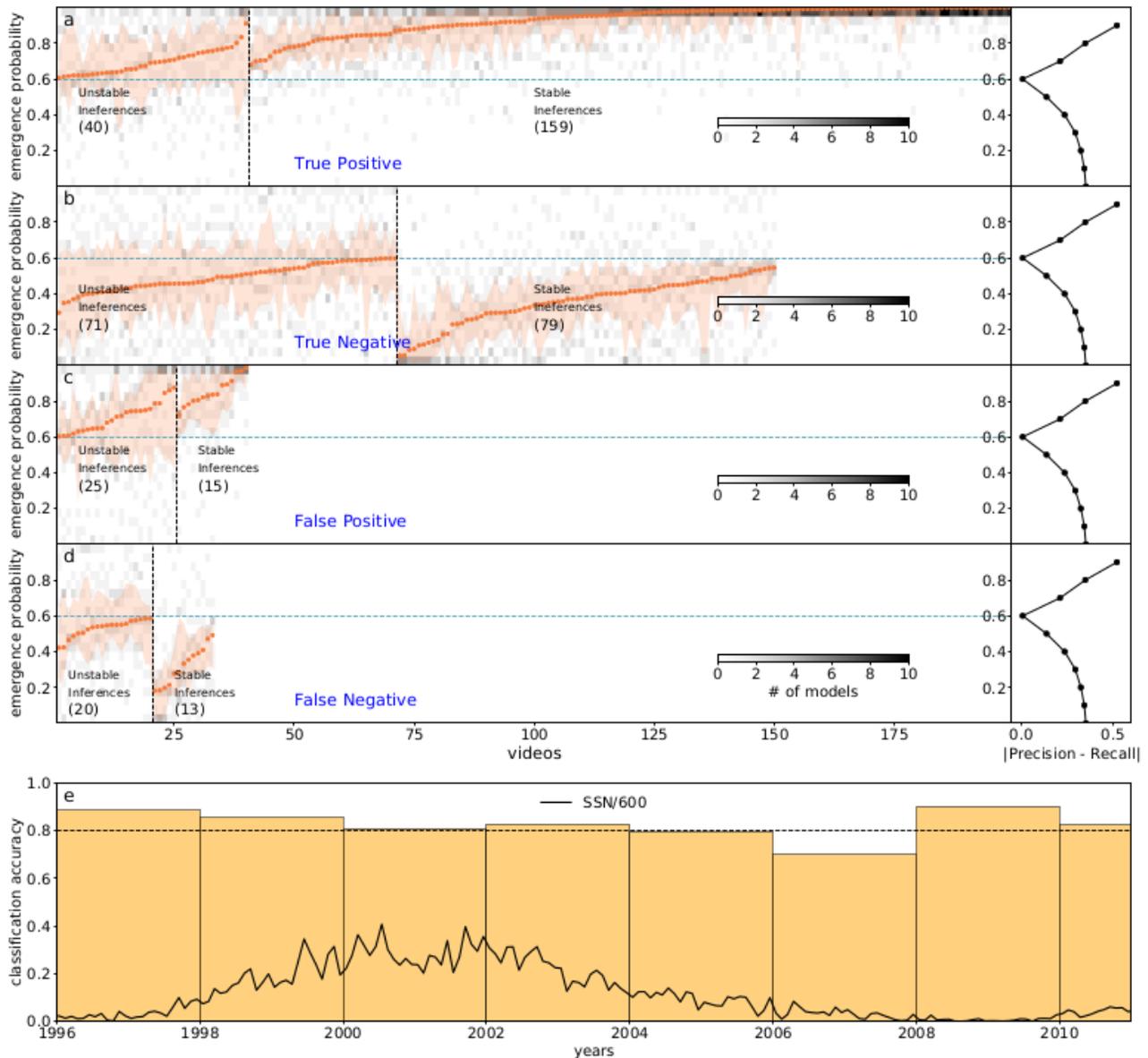

**Figure 4.** Model ensemble used to estimate the classification accuracy on test set. Model inferences are depicted as 2D histograms in the panels a-d. The x-axis indicates video number and the y-axis shows sigmoid output of the CNN-ensemble. The gray scale depicts the number of CNNs with a certain sigmoid probability output for a particular video. The median of model-ensemble outputs for a given video is marked with orange dots and the videos are arranged in ascending order of medians. Videos are classified as 'emergence (positive)' or 'non-emergence (negative)' depending on whether the median probability exceeds a threshold of 0.6 (dashed line) that is found as the intersection point of the precision and recall curves as shown by the black curves on the right. The shaded regions enclose the $25^{th}$ and $75^{th}$ percentiles of the distribution of model outputs for each video. To estimate the uncertainty in classification accuracy, the ensemble inferences are divided into stable, unstable ones as shown in all the panels separated by a vertical dashed line. For stable (unstable) inferences, the probability threshold of 0.6 lies outside (inside) the ensemble $25^{th}$-$75^{th}$ percentile range. Panel e shows the classification accuracy, marked by heights of orange bars, evaluated on the test set as a function of solar cycle phase. The dashed horizontal line marks a classification accuracy of 80% as reference. The monthly mean total sunspot number from SILSO (publicly available at https://wwwbis.sidc.be/silso/) is depicted with the solid black curve to show solar cycle phase.

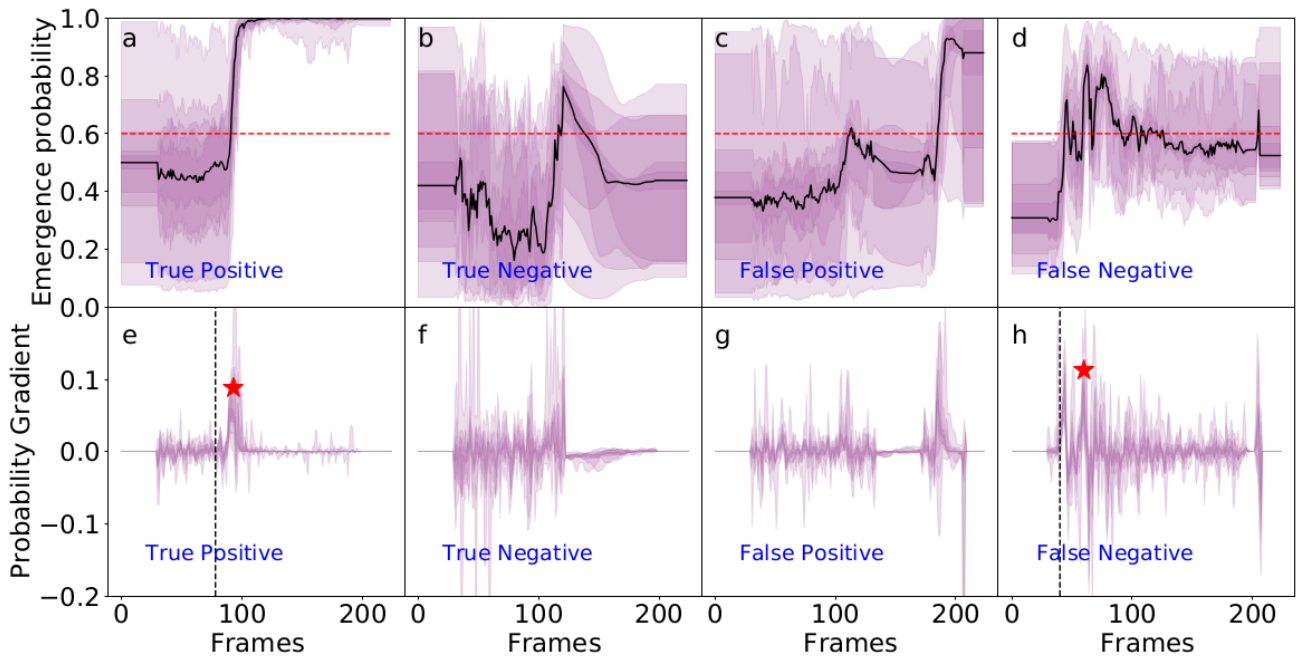

**Figure 5.** Identifying emergence epoch by frame stacking. Panels a-d show how the emergence probability changes for a video when frame truncation is gradually reduced to cover all the original frames and subsequently videos are classified as positive or negative with a threshold of 0.6 depicted by the red horizontal dashed line. Panels e-h show the identification of emergence epoch (depicted by a red star) locating the frame of maximum emergence probability gradient. Top left panel depicts the same event as shown in Figure 1. The ground truth frames are marked by the vertical dashed lines. To show the central tendency of probability and probability gradient the region values for each frame number are sorted and regions between equidistant points from extrema are shaded.

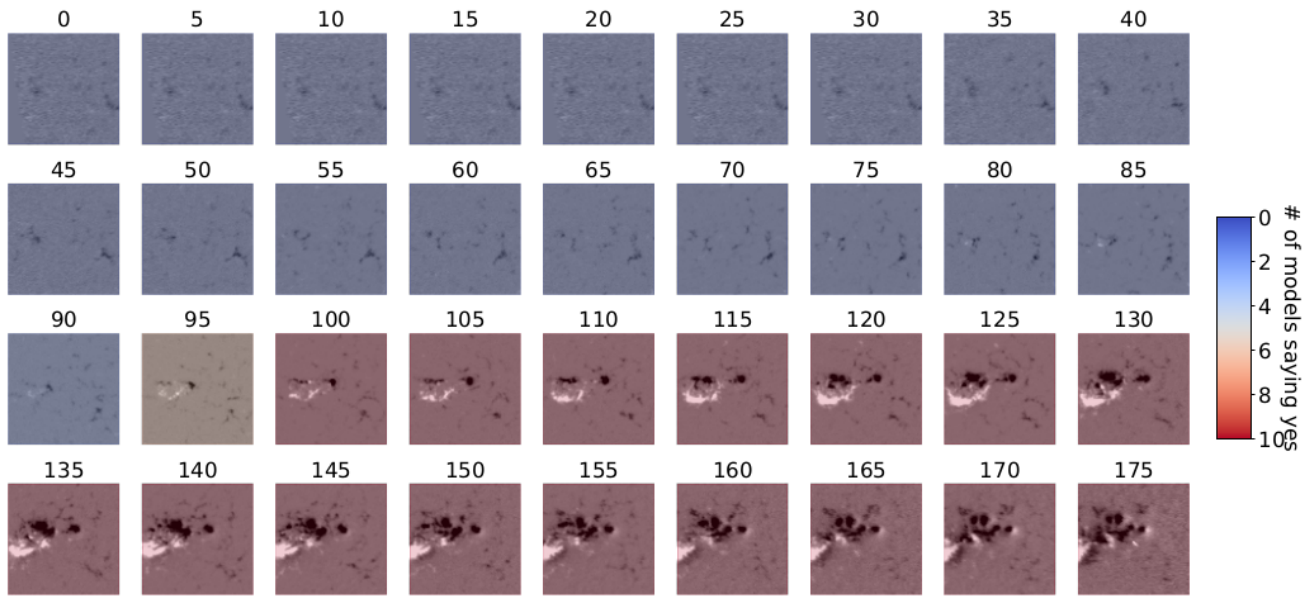

**Figure 6.** Emergence epoch identification using an ensemble of models. The panels correspond to every 5th frame of the 'bipolar emergence' video also shown in Figure 1. Each frame is color-coded according to the number of models that classify the video as 'emergence' until that frame.

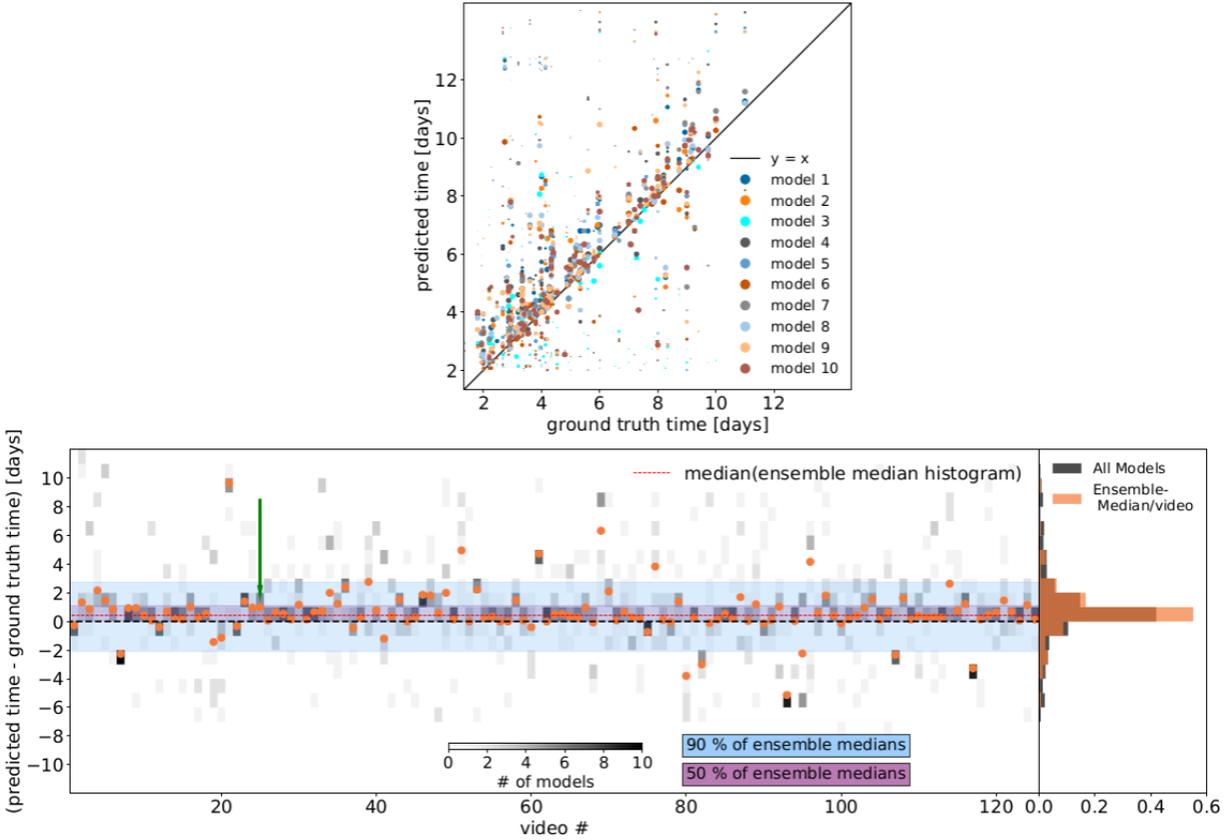

**Figure 7.** Accuracy of emergence epoch identification. Top panel shows predicted time of emergence vs ground-truth time for each ensemble member. The size of the symbols are scaled according to their distance (D) from ensemble-mean (symbol size = $10 e^{-\left(\frac{D}{\sigma}\right)^2}$; σ is the ensemble-standard deviation) for each video on the test set. Bottom panel shows a 2D histogram representing the joint distribution of the 'bipolar emergence' videos and predicted emergence time w.r.t. ground truth using model ensemble. The gray level represents the number of models predicting within a time bin (1 day) for a particular video. The green arrow points to the location of the video in top panel in the 2D histogram. Median time for over the models for each video is depicted by orange dot. The ground truth reference is depicted by the black dashed horizontal line. The histograms on the right show occurrence of prediction over all models (black) and model ensemble medians per video (orange). The median of the orange histogram is depicted by the dashed orange horizontal line. It clearly shows that the models have a tendency to detect the emergence little later (≈0.4 day) than the actual initiation (also seen in Figure 5). The purple and light blue shaded region show the time range within which $25^{th}$-$75^{th}$ percentile and $5^{th}$-$95^{th}$ percentile of predictions lie respectively.